# Computing in Covert Domain Using Data Hiding

Zhenxing Qian, Zichi Wang, and Xinpeng Zhang

*Abstract*—This paper proposes an idea of data computing in the covert domain (DCCD). We show that with information hiding some data computing tasks can be executed beneath the covers like images, audios, random data, etc. In the proposed framework, a sender hides his source data into two covers and uploads them onto a server. The server executes computation within the stego and returns the covert computing result to a receiver. With the covert result, the receiver can extract the computing result of the source data. During the process, it is imperceptible for the server and the adversaries to obtain the source data as they are hidden in the cover. The transmission can be done over public channels. Meanwhile, since the computation is realized in the covert domain, the cloud cannot obtain the knowledge of the computing result. Therefore, the proposed idea is useful for cloud computing.

## I. INTRODUCTION

With the development of big data and cloud computing, many security problems appear, e.g., privacy disclosure [1], data abuse [2], malicious attacks [3]. Therefore, many data protection algorithms have been proposed for data storage and computing. One important task of protection is to conceal the content of sensitive data. For example, during cloud computing the users always hope that the cloud can process the committed data without knowing its content [4]. Data encryption is one of the most effective and popular means of privacy protection [5]. However, protection by encryption would inevitably impact the utility of cloud computing.

As present, the most popular approaches to achieve secure computation for cloud is signal processing in encrypted domain [6-9]. As shown in Fig. 1, a sender encrypts his source data (the sensitive information that cannot be exposed to the others), and uploads the encrypted versions to the cloud sever. The server executes computing within the encrypted data without knowing the content of source data, and returns the computing results to a receiver. Finally, the receiver decrypts the returned results to obtain the computation results the of source data. Homomorphic encryption algorithms were proposed to achieve these goals. After encrypting the data into a ciphertext with tremendous bits, additive or multiplicative operations can be done in encrypted domain. As the computation complexity [10-12] is very large, homomorphic encryption is not convenient for common users or mobile devices.

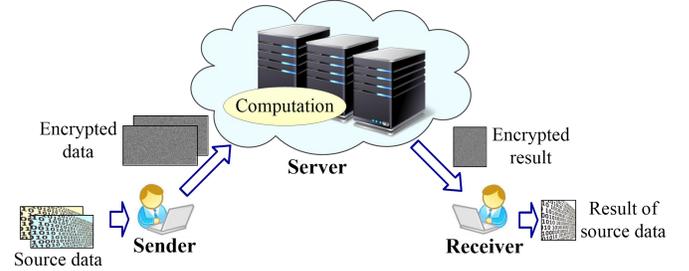

Fig. 1 Computation in encrypted domain using encryption

In this paper, we propose a novel idea of secure computation, i.e., data computing in in the covert domain (**DCCD**). Different from the encrypted domain, the covert domain is a space generated by information hiding that can accommodate data inside a cover. The covers can be images, videos, audios, noise, or even random data, etc. Traditionally, information hiding is also called steganography, which is used to transmit secret message covertly over public channels. Many works have been done on steganography, e.g., LSB, JSteg, ZZW, STC, etc. We find that information hiding can also be used for cloud computing. On the one hand, data privacy can be protected since the data are embedded inside the covers. On the other hand, both the additive and multiplicative operations can be realized in the covert data. As shown in Fig. 2, A sender embeds his source data into covers, and then uploads the obtained stegos to a server. For the ease of illustration we use images as covers for example. The server executes computing within the covers and returns the stego with computing result to the receiver. The receiver extracts the computation result of the source data. In this framework, no information of the original data can be obtained by the server. Meanwhile, the computation complexity is much smaller than the homomorphic encryption. Besides, the source data are secure to be transmitted over public channels since the data are hidden inside the covers.

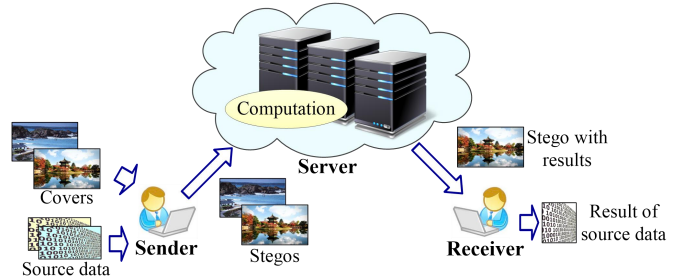

Fig. 2 Computation in covert domain using information hiding

This work was supported by the Natural Science Foundation of China (Grant U1736213).

Zhenxing Qian is with Shanghai Institute of Intelligent Electronics and Systems, School of Computer Science, Fudan University, Shanghai, 201203, P.R., China. Zichi Wang and Xinpeng Zhang are with Shanghai Institute for Advanced Communication and Data Science, Key laboratory of Specialty Fiber Optics and Optical Access Networks, Joint International Research Laboratory of Specialty Fiber Optics and Advanced Communication, Shanghai University, Shanghai, 200444, P.R., China. (Corresponding author: Zhenxing Qian, E-mail: zxqian@fudan.edu.cn)

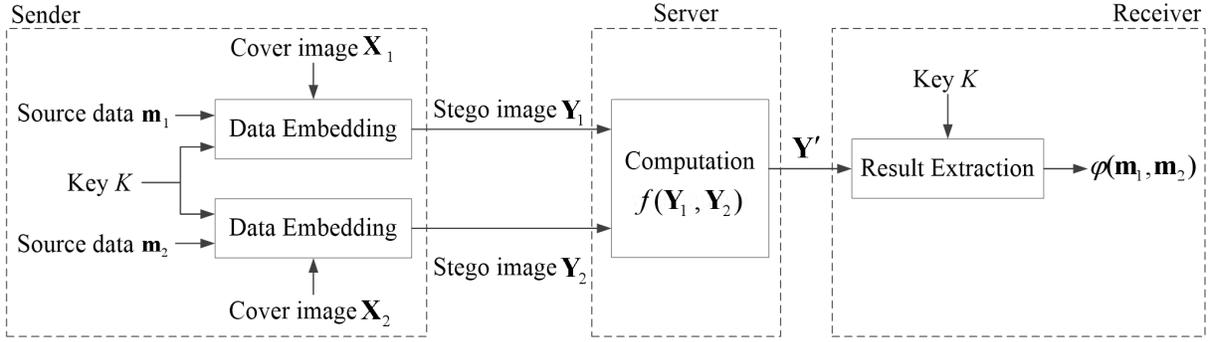

Fig. 3 Framework of the proposed method.

## II. PROPOSED METHOD

The proposed framework of DCCD is shown in Fig. 3. We use digital images as covers to show the process of DCCD. A sender hopes to calculate $\varphi(\mathbf{m}_1,\mathbf{m}_2)$ in the server, while keep $\mathbf{m}_1$ and $\mathbf{m}_2$ secret to the server, where $\mathbf{m}_1$ and $\mathbf{m}_2$ are the source data, and $\varphi(\cdot)$ is a calculation operator.

To achieve this goal, the sender embeds $\mathbf{m}_1$ and $\mathbf{m}_2$ into the cover images $\mathbf{X}_1$ and $\mathbf{X}_2$ with data hiding key $K$, respectively. The stego images $\mathbf{Y}_1$ and $\mathbf{Y}_2$ are then uploaded onto the server. The server executes the computation $f(\mathbf{Y}_1,\mathbf{Y}_2)$ using $\mathbf{Y}_1$ and $\mathbf{Y}_2$ without any knowledge of $\mathbf{m}_1$ and $\mathbf{m}_2$. After that, the stego $\mathbf{Y}'$ containing computation result is obtained. With $\mathbf{Y}'$ the receiver is able to extract the computation result $\varphi(\mathbf{m}_1, \mathbf{m}_2)$.

The computation is done in the covert domain generated by data hiding. Many algorithms can be used to realize the proposed DCCD. For the ease of explanation, we use the matrix embedding in LSB (Least Significant Bitplane) as examples to illustrate the proposed idea.

### A. Source data embedding

Let the source data be binary sequences, i.e., $\mathbf{m}_1=[m_1(1), m_1(2), \ldots, m_1(k)]^T \in \{0,1\}^{k \times 1}$, $\mathbf{m}_2=[m_2(1), m_2(2), \ldots, m_2(k)]^T \in \{0,1\}^{k \times 1}$. To conceal $\mathbf{m}_1$ and $\mathbf{m}_2$, two cover images $\mathbf{X}_1=[x_1(i,j)]^{w \times r}$ and $\mathbf{X}_2=[x_2(i,j)]^{w \times r}$ is used, where $i \in \{1, 2, \ldots, w\}, j \in \{1, 2, \ldots, r\}$. The LSB $\mathbf{c}_1=[c_1(i,j)] \in \{0,1\}^{w \times r}$ and $\mathbf{c}_2=[c_2(i,j)] \in \{0,1\}^{w \times r}$ of $\mathbf{X}_1$ and $\mathbf{X}_2$ are used for embedding, as shown in (1) and (2).

$$c_1(i,j) = \mathrm{mod}(x_1(i,j), 2) \quad (1)$$

$$c_2(i,j) = \mathrm{mod}(x_2(i,j), 2) \quad (2)$$

Next, we use matrix embedding method [13] to embed $\mathbf{m}_1$ into $\mathbf{c}_1$. With a data hiding key $K$, we generate an binary matrix $\mathbf{H}=[h(u,v)] \in \{0,1\}^{k \times wr}$, where $u \in \{1, 2, \ldots, k\}, v \in \{1, 2, \ldots, wr\}$. The data hiding key is shared by the sender and the receiver. The process of data embedding is to make (3) true by modifying the elements in $\mathbf{c}_1$ to generate $\hat{\mathbf{c}}_1$.

$$\mathbf{m}_1 = \mathbf{H}\hat{\mathbf{c}}_1 \quad (3)$$

During the embedding, $\hat{\mathbf{c}}_1 = [\hat{c}_1(1), \hat{c}_1(2), \ldots, \hat{c}_1(wr)]^T \in \{0,1\}^{wr \times 1}$ is obtained by cascading the rows in $\mathbf{c}_1$, as shown in (4), $t \in \{1, 2, \ldots, wr\}$ and "$\lfloor \cdot \rfloor$" is the floor rounding operator.

$$\hat{c}_1(t) = c_1\left(\left\lfloor \frac{t-1}{w} \right\rfloor + 1, t - \left\lfloor \frac{t-1}{w} \right\rfloor w\right) \quad (4)$$

Subsequently, we substitute the LSB $\mathbf{c}_1$ in $\mathbf{X}_1$ to achieve the stego $\mathbf{Y}_1$.

The procedure of embedding $\mathbf{m}_2$ into $\mathbf{c}_2$ is the same as embedding $\mathbf{m}_1$ into $\mathbf{c}_1$. After cascading the rows in $\mathbf{c}_2$ to obtain $\hat{\mathbf{c}}_2 = [\hat{c}_2(1), \hat{c}_2(2), \ldots, \hat{c}_2(wr)]^T \in \{0,1\}^{wr \times 1}$, we make (5) true by modifying the elements in $\mathbf{c}_2$ to achieve $\hat{\mathbf{c}}_2$. Therefore, the stego $\mathbf{Y}_2$ is obtained.

$$\mathbf{m}_2 = \mathbf{H}\hat{\mathbf{c}}_2 \quad (5)$$

After the data embedding, we upload the stego images $\mathbf{Y}_1$ and $\mathbf{Y}_2$ to the server. Without the data hiding key, the content of $\mathbf{m}_1$ and $\mathbf{m}_2$ cannot be revealed. This security performance will be demonstrated in *Subsection III. C*.

### B. Computation in Covert Domain

With the stego images $\mathbf{Y}_1$ and $\mathbf{Y}_2$, the server can execute the computation task $f(\mathbf{Y}_1,\mathbf{Y}_2)$, which is equivalent to $\varphi(\mathbf{m}_1,\mathbf{m}_2)$ in the plaintext domain. Next, we discuss three cases of binary calculations $\varphi(\cdot)$ in the covert domain, i.e.,

$$\varphi_1(\mathbf{m}_1, \mathbf{m}_2) = \mathbf{m}_1 + \mathbf{m}_2,$$

$$\varphi_2(\mathbf{m}_1, \mathbf{m}_2) = \mathbf{m}_1\mathbf{m}_2^T,$$

$$\varphi_3(\mathbf{m}_1, \mathbf{m}_2) = \mathbf{m}_1^T\mathbf{m}_2.$$

These operations are widely used outsourcing computation [14,15], image retrieval [16,17], privacy protection [18,19], etc.

*i) Case One*

For the case of $\varphi_1(\mathbf{m}_1,\mathbf{m}_2)$, we use

$$f(\mathbf{Y}_1,\mathbf{Y}_2) = \hat{\mathbf{c}}_1 + \hat{\mathbf{c}}_2 \quad (6)$$

In this way, the LSB $\mathbf{c}' = [c'(i,j)] \in \{0,1\}^{w \times r}$ of $\mathbf{Y}'$ is

$$c'(i,j) = y_1(i,j) + y_2(i,j) \quad (7)$$

After cascading, the obtained binary sequence $\hat{\mathbf{c}}' = [\hat{c}'(1), \hat{c}'(2), \ldots, \hat{c}'(wr)]^T \in \{0,1\}^{wr \times 1}$ is

$$\hat{c}'(t) = c'\left(\left\lfloor \frac{t-1}{w} \right\rfloor + 1, t - \left\lfloor \frac{t-1}{w} \right\rfloor w\right) \quad (8)$$

which satisfies

$$\hat{\mathbf{c}}' = \hat{\mathbf{c}}_1 + \hat{\mathbf{c}}_2 \quad (9)$$

After that, the server replaces the LSB of $\mathbf{Y}_1$ by $\hat{\mathbf{c}}'$ to generate the final $\mathbf{Y}'$, which is sent to the receiver.

According to (3) and (5), the receiver can obtain

$$\mathbf{H}\hat{\mathbf{c}}' = \mathbf{H}\hat{\mathbf{c}}_1 + \mathbf{H}\hat{\mathbf{c}}_2 = \mathbf{m}_1 + \mathbf{m}_2 \quad (10)$$

Therefore, the result extracted from $\mathbf{Y}'$ is $\varphi_1(\mathbf{m}_1, \mathbf{m}_2) = \mathbf{m}_1 + \mathbf{m}_2$.

*ii) Case Two*

For the case of $\varphi_2(\mathbf{m}_1, \mathbf{m}_2)$, we calculate

$$f(\mathbf{Y}_1, \mathbf{Y}_2) = \hat{\mathbf{c}}' = \hat{\mathbf{c}}_1 \hat{\mathbf{c}}_2^\mathrm{T},$$

and put back the results into the LSB of $\mathbf{Y}_1$ to generate $\mathbf{Y}'$.

The receiver can obtain the result of $\mathbf{m}_1 \mathbf{m}_2^\mathrm{T}$ by $\mathbf{H}\hat{\mathbf{c}}'\mathbf{H}^\mathrm{T}$ because of

$$\mathbf{m}_1 \mathbf{m}_2^\mathrm{T} = \mathbf{H}\hat{\mathbf{c}}_1 (\mathbf{H}\hat{\mathbf{c}}_2)^\mathrm{T} = \mathbf{H}\hat{\mathbf{c}}_1 \hat{\mathbf{c}}_2^\mathrm{T} \mathbf{H}^\mathrm{T} = \mathbf{H}\hat{\mathbf{c}}'\mathbf{H}^\mathrm{T}. \quad (11)$$

*iii) Case Three*

For the case of $\varphi_3(\mathbf{m}_1, \mathbf{m}_2) = \mathbf{m}_1^\mathrm{T} \mathbf{m}_2$, we use

$$f(\mathbf{Y}_1, \mathbf{Y}_2) = \hat{\mathbf{c}}' = \hat{\mathbf{c}}_1^\mathrm{T} \hat{\mathbf{c}}_2.$$

The server calculates

$$\mathbf{m}_1^\mathrm{T} \mathbf{m}_2 = (\mathbf{H}\hat{\mathbf{c}}_1)^\mathrm{T} \mathbf{H}\hat{\mathbf{c}}_2 = \hat{\mathbf{c}}_1^\mathrm{T} \mathbf{H}^\mathrm{T} \mathbf{H}\hat{\mathbf{c}}_2. \quad (12)$$

Once the embedding matrix $\mathbf{H}$ is an orthogonal matrix, i.e.,

$$\mathbf{H}^\mathrm{T} \cdot \mathbf{H} = \mathbf{I}$$

The result of (12) is equal to

$$\mathbf{m}_1^\mathrm{T} \mathbf{m}_2 = \hat{\mathbf{c}}' = \hat{\mathbf{c}}_1^\mathrm{T} \hat{\mathbf{c}}_2.$$

After that, the server replaces the LSB of $\mathbf{Y}_1$ by $\hat{\mathbf{c}}'$ to generate the final $\mathbf{Y}'$.

As $\hat{\mathbf{c}}'$ is equal to $\mathbf{m}_1^\mathrm{T} \mathbf{m}_2$, the server can obtain the calculation result. In order to protect the result, the sender can multiply the source data $\mathbf{m}_1$ and $\mathbf{m}_2$ with two factors, respectively. On the other side, the receiver can remove these factors to obtain the real result. Therefore, the security of data computation can be guaranteed.

## III. EXPERIMENTAL RESULTS

We have conducted many experiments to verify the DCCD idea. In this section, we show the feasibility, imperceptibility, security performance, and computational complexity of the proposed paradigm.

*A. Feasibility*

We use two random binary sequences with 1000 bits are used as $\mathbf{m}_1$ and $\mathbf{m}_2$. As shown in Fig. 4, the test images Lena and Baboon sized 512×512 are used as the cover $\mathbf{X}_1$ and $\mathbf{X}_2$. After data embedding, $\mathbf{m}_1$ and $\mathbf{m}_2$ are embedded into $\mathbf{X}_1$ and $\mathbf{X}_2$ to generate the stego images $\mathbf{Y}_1$ and $\mathbf{Y}_2$, as shown in Fig. 5. The stego images are indistinguishable from cover images.

Therefore, the imperceptibility of the source data can be realized. More analysis about the imperceptibility of $\mathbf{m}_1$ and $\mathbf{m}_2$ are shown in Fig. 6.

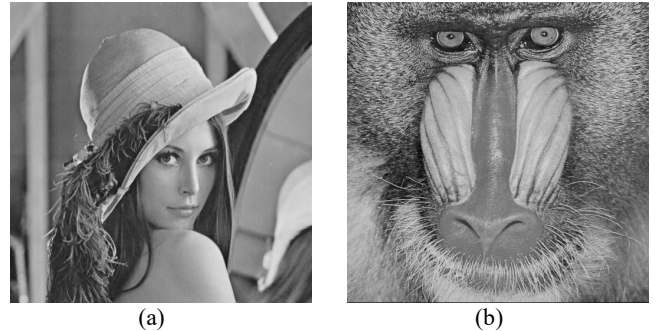

(a)      (b)
Fig. 4 Cover images (a) Lena; (b) Baboon.

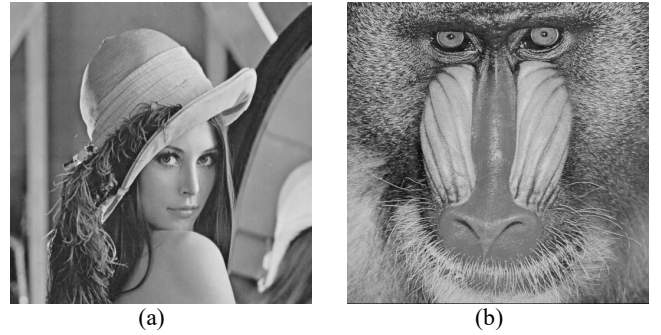

(a)      (b)
Fig. 5 Stego images (a) Lena containing $\mathbf{m}_1$; (b) Baboon containing $\mathbf{m}_2$.

Next, we process DCCD with $\mathbf{Y}_1$ and $\mathbf{Y}_2$ using the algorithms described in subsection II.B. The result extract form the covert domain are then compared with $\varphi(\mathbf{m}_1, \mathbf{m}_2)$. The difference ratio between $\varphi(\mathbf{m}_1, \mathbf{m}_2)$ and the extracted result are shown in Table 1. For all cases, the difference ratio is zero. It means the receiver can obtain $\varphi(\mathbf{m}_1, \mathbf{m}_2)$ after data extraction. Hence, the proposed method is feasible.

Table 1. Difference ratio between $\varphi(\mathbf{m}_1, \mathbf{m}_2)$ and the extracted result

| $\varphi(\mathbf{m}_1, \mathbf{m}_2)$ | $\mathbf{m}_1 + \mathbf{m}_2$ | $\mathbf{m}_1 \mathbf{m}_2^\mathrm{T}$ | $\mathbf{m}_1^\mathrm{T} \mathbf{m}_2$ |
|---|---|---|---|
| Difference ratio | 0 | 0 | 0 |

*B. Imperceptibility*

To verify the imperceptibility of the source data statistically, we use all 1338 images sized 512×384 in UCID [20] as cover images. Each image is embedded with capacity 1000, 2000, 3000, 4000, and 5000 bits respectively using the proposed method. The imperceptibility of the source data can be checked by modern steganalytic methods which are based on the supervised machine learning [21]. Specifically, we employ the popular steganalytic feature extraction methods SPAM [22], SRMQ1 [23], SRM [23], and PSRM [24] with an ensemble classifier [25]. One half of the cover and stego feature sets are used for training, while the remaining sets are used for testing. The criterion of evaluating the performance of feature sets is the minimal total error $P_\mathrm{E}$ with identical priors achieved on the testing sets [25].

$$P_E = \min_{P_{FA}} \left( \frac{P_{FA} + P_{MD}}{2} \right) \quad (13)$$

where $P_{FA}$ is the false alarm rate and $P_{MD}$ the missed detection rate. The performance is evaluated using the average of $P_E$ over ten random tests. A higher $P_E$ value stands for a better imperceptibility.

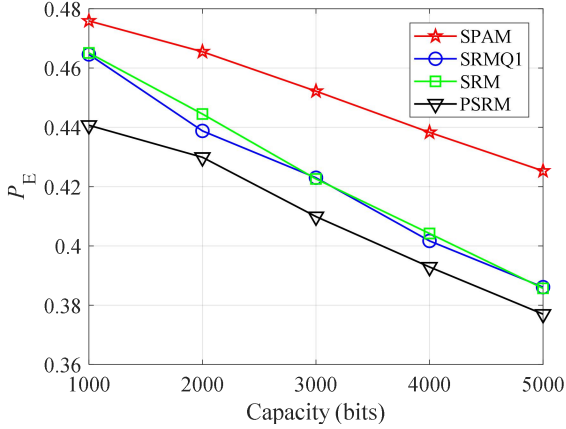

Fig. 6 Imperceptibility of data embedding.

The imperceptibility of data embedding is shown in Fig. 6. The values of $P_E$ keep a high level for all cases. It is difficult to discover the embedded source data from the stego images. Thus, the imperceptibility can be guaranteed. In addition, the values of $P_E$ are close to 0.5 (the bound of $P_E$) for small capacity (such than 1000 bits). The result indicates that the existence of source data is completely undetectable using the modern steganalysis tools when capacity is less than 1000 bits. Therefore, the specific attention and unwanted attack existing in encrypted domain can be avoided in the covert domain.

*C. Security performance*

In the proposed framework of DCCD, the content of source data should not be leaked to the server. In other words, the content of $\mathbf{m}_1$ and $\mathbf{m}_2$ cannot be obtained without data hiding key. To verify the security of source data, we also use the images in UCID as covers. Each image is embedded with capacity 1000, 2000, 3000, 4000, and 5000 bits, respectively. Next, a binary matrix with random bits (without data hiding key) is generated for each stego image to extract source data. Meanwhile, a binary matrix generated using data hiding key is used for data extraction. The average ratio of data extraction error is shown in Table 2.

Table 2. Data extraction error with/without data hiding key.

| Capacity (bits) | 1000 | 2000 | 3000 | 4000 | 5000 |
|---|---|---|---|---|---|
| Error with data hiding key (%) | 0 | 0 | 0 | 0 | 0 |
| Error without data hiding key (%) | 50.02 | 49.96 | 49.97 | 49.96 | 49.95 |

The result show that the data extraction error is around 50% (the bound of extraction error) for all cases. It means the content of source data would not be leaked without the data hiding key. On the other hand, the source data can be extracted correctly (data extraction error is 0%) when data hiding key is known. Therefore, the security of the proposed method can be guaranteed.

*D. Computational complexity*

While the complexity of computation in the encrypted domain is huge, the complexity of computing in the covert domain is smaller. To verify the efficiency of the proposed DCCD scheme, source data with 1000, 2000, 3000, 4000, and 5000 bits are embedded into image Lena, respectively. Meanwhile, the same source data is encrypted using RSA with the key sized 256 bits. The complexity comparison between the proposed DCCD and the RSA is shown in Fig. 7. These results are generated on a serve with 1.8 GHz CPU, 8 GB memory and windows 10. The type of system is 64 bits and the version of MATLAB is R2017b.

The result show that the computational complexity of DCCM is much smaller than that of RSA for all cases. In addition, the computational complexity of RSA increases when a longer key is used. Thus, the proposed DCCD idea is more convenient for cloud computing.

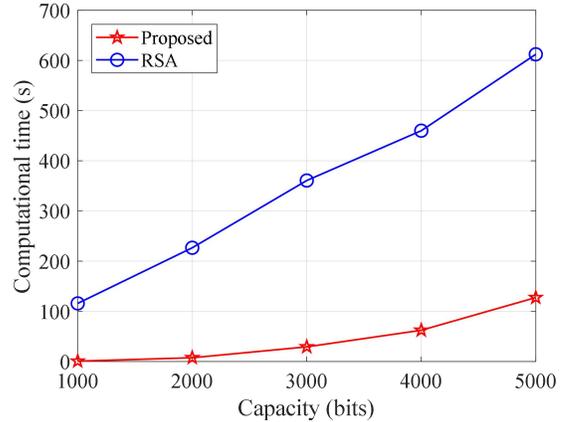

Fig. 7 computational complexity comparison.

IV. CONCLUSIONS

We propose a new idea of achieving secure computation in the covert domain using data hiding. A practical method is designed to implement the binary additive and multiplication calculations. As the source data are hidden in the covers, it is imperceptible for the cloud and the adversaries to obtain the data. Meanwhile, as the computation is realized in the covert domain, the cloud cannot obtain the computation data and the result. On the recipient side, the computation result can be obtained after data extraction. Experimental results show the effectiveness of the proposed DCCD idea. To the best of our knowledge, it is the first work on computing in covert domain. More works on this topic can be done in the future.